\documentclass[prb,aps,twocolumn,groupedaddress,showpacs,floatfix]{revtex4-1}

\usepackage[latin1]{inputenc}
\usepackage{amsfonts}
\usepackage{amsmath,amssymb}
\usepackage{graphicx,pst-plot,pst-node}
\usepackage{bm}
\usepackage{subfigure}
\usepackage{psfrag}
\usepackage{hyperref}
\usepackage[mathscr]{eucal} 
\usepackage{bbm}
\usepackage{xcolor}

\newcommand{\ec}{\frac{e^2}{\epsilon\ell}}

\newcommand{\ZZ}{{\mathbb Z}}
\newcommand{\sgn}{\text{sgn}}

\newcommand{\half}{\frac{1}{2}}
\newcommand{\td}{\text{d}}
\newcommand{\ti}{\text{i}}

\newcommand{\Ne}{N_{\text{e}}}

\newcommand{\cdg}{c^{\dagger}}

\newcommand{\vq}{\vec{q}}
\newcommand{\vcr}{\vec{r}}

\newcommand{\vL}{\vec{L}}

\newcommand{\Nf}{N_{\phi}}

\newcommand{\Ltx}{L_{2x}}
\newcommand{\Lty}{L_{2y}}

\newcommand{\tn}{\tilde{n}}

\newcommand{\vk}{\vec{k}}

\newcommand{\vQ}{\vec{Q}}

\newcommand{\vQp}{\vQ_{\parallel}}
\newcommand{\Qp}{Q_{\parallel}}
\newcommand{\scF}{\mathcal{F}}

\newcommand{\nLL}{n_{\text{LL}}}
\newcommand{\bC}{\bar{C}}
\newcommand{\nLLa}{n_{\text{LL}1}}
\newcommand{\nLLb}{n_{\text{LL}2}}
\newcommand{\nLLc}{n_{\text{LL}3}}
\newcommand{\nLLd}{n_{\text{LL}4}}

\newcommand{\Bc}{B_{\text{c}}}

\begin{document}
\title {Fractional Quantum Hall Effects in HgTe Quantum Wells}

\author{Jianhui Wang}
\affiliation{Department of Physics, Ben-Gurion University of the Negev, Beer Sheva 84105, Israel}
\affiliation{Department of Condensed Matter Physics, Weizmann Institute of Science, Rehovot 76100, Israel}

\begin{abstract}
We study the possibility of fractional quantum Hall effects in HgTe quantum wells using exact diagonalization. Our results show that Laughlin states, the Moore-Read state, and the Read-Rezayi $Z_3$ state can all be supported. However, near the level crossing point (of the single-particle spectrum), the gap can be destroyed by Landau level mixing, and the Moore-Read state and the Read-Rezayi state dominate over their respective competing states only for wide wells. For smaller well widths the Moore-Read state crosses over to the composite fermion Fermi sea, while the Read-Rezayi state loses its dominance over the hierarchy state.
\end{abstract}

\date{\today}

\maketitle

HgTe quantum wells have attracted a lot of attentions in the last several years. This is mainly due to the prediction \cite{BHZ} and observation \cite{Molenkamp} of the quantum spin Hall effect in this type of systems. However, it is natural to ask whether it can exhibit other phenomena associated with two-dimensional (2D) electronic systems. One of the most prominent such phenomena is the fractional quantum Hall effect. This phenomenon is interesting and important due to its strongly correlated nature and possible application in topological quantum computation. Fractional quantum Hall effect in HgTe quantum well is interesting because of the strong spin-orbit interaction in this material. The effect of strong spin-orbit interaction is two-fold. First, it affects the single-particle wavefunctions, so even though the electron-electron interaction is the Coulomb interaction just like in other 2D materials, the matrix elements of the Coulomb interaction are different. This is equivalent to electrons with a parabolic band interacting with an effective interaction that is different from the Coulomb interaction. This is an important point because at some filling factors there are competitions between different incompressible states, even between incompressible states and compressible states, and which state is realized and how stable this state is depend on the details of the interaction. Second, the single-particle Landau level spectrum is also modified by the strong spin-orbit interaction. In particular, there are values of the magnetic field for which two Landau levels (LLs) cross each other. LL mixing could become important around such points.

In this work we study the possibilities of different fractional quantum Hall effects in HgTe quantum well using exact diagonalization in the torus geometry\cite{Yoshioka1,Su,Haldane,BR}. The effect of spin-orbit interaction on the single-particle wavefunction is taken into account with an eight-band calculation of the envelope function \cite{Wuerzburg-band-structure}. In the presence of a magnetic field, the envelope function has, with the axial approximation, the form \begin{equation}\psi_{\nLL,k_x}(\vcr)=\exp(\text{i}k_x x)\left(\begin{array}{cc} f_1^{(\nLL)}(z) & \varphi_{\nLL} \\ f_2^{(\nLL)}(z) & \varphi_{\nLL+1} \\ f_3^{(\nLL)}(z) & \varphi_{\nLL-1} \\ f_4^{(\nLL)}(z) & \varphi_{\nLL} \\ f_5^{(\nLL)}(z) & \varphi_{\nLL+1} \\ f_6^{(\nLL)}(z) & \varphi_{\nLL+2} \\ f_7^{(\nLL)}(z) & \varphi_{\nLL} \\ f_8^{(\nLL)}(z) & \varphi_{\nLL+1}\end{array}\right),\label{envelop-func}
\end{equation}
where $k_x$ is the single-particle momentum in the $x$ direction [the difference in the prefactor from Ref.~\onlinecite{Wuerzburg-band-structure} is due to a different choice of the vector potential, $\vec{A}=-By\hat{x}$, with $\hat{x}$ being the unit vector in the $x$-direction], and $e^{\text{i}k_x x}\varphi_n$ are Landau wavefunctions for a parabolic band. By a parabolic band we mean a (strictly 2D) system with parabolic dispersion in the absence of a magnetic field, i.e. $H_{\text{para}}=\frac{\vec{p}^{\,2}}{2m}$ [$\vec{p}=(p_x,p_y)$ being the 2D momentum], which becomes $H_{\text{para,mag}}=\frac{1}{2m}(\vec{p}+\frac{e}{c}\vec{A})^2$ in the presence of a magnetic field. The solution of the parabolic problem is standard\cite{Jain-book} and $\varphi_n=(-1)^n[\pi\ell^2 2^{2n}(n!)^2]^{-\frac{1}{4}} e^{-\half (\frac{y}{\ell}-\ell k_x)^2}H_n(\frac{y}{\ell}-\ell k_x)$, where $\ell$ is the magnetic length and $H_n$ is the Hermite polynomial. $\varphi_n$ appear in the ansartz (\ref{envelop-func}) because they form a basis for the raising and lowering operators, in terms of which the $p_y$ operator can be expressed. [The phase $(-1)^n$ is chosen so that we have $a\varphi_n=\sqrt{n}\varphi_{n-1}$ and $a^{\dagger}\varphi_n=\sqrt{n+1}\varphi_{n+1}$ with the correct sign, when the raising and lowering operators are defined as in Ref.~\onlinecite{Wuerzburg-band-structure}, which in the current gauge have the explicit forms $a=\frac{1}{\sqrt{2}}(k_x\ell-\frac{y}{\ell}-\partial_y), a^{\dagger}=\frac{1}{\sqrt{2}}(k_x\ell-\frac{y}{\ell}+\partial_y)$.] Because of the specific forms of the equations for the envelope functions in the axial approximation, with the choice of the indices of the $\varphi_n$ as in Eq.~(\ref{envelop-func}), the $y$-dependence of each equation reduces to a common factor of $\varphi_n$, leaving us a system of ordinary differential equations for the $f_i(z)$, which is then solved numerically.

$\varphi_n$ with $n<0$ are understood to be 0, so the lowest $\nLL$ is $-2$, and $\psi_{-2}$ has only one nonzero component, containing $\varphi_0$. (The negative sign of the index in $\psi_{-2}$ is of no significance; one could have shifted the origin of $\nLL$ by defining $\nLL'\equiv \nLL+2$, then $\nLL'$ would take nonnegative integer values. We are simply following the convention used in Ref.~\onlinecite{Wuerzburg-band-structure}.) Hence $\psi_{-2}$ is similar to the lowest LL in systems with parabolic bands, in the sense that the only nonzero component of Eq.~(\ref{envelop-func}) is proportional to $\varphi_0$, except that here there is a finite width in the $z$ direction, i.e. there is an additional factor $f_6^{(0)}(z)$ (there is no such factor for strictly 2D electron gas). To avoid possible confusion, we repeat that the word ``parabolic'' refers to the dispersion in the absence of a magnetic field for a system whose eigenfunctions [$e^{\text{i}k_x x}\varphi_n(y)$] in the presence of a magnetic field we use as building blocks in Eq.~(\ref{envelop-func}). This parabolic case is also the usual approximation for 2D electron gas. However, there is nothing parabolic in our current case otherwise. In particular, the confinement in the current case is due to the discontinuity of the conduction and valence band edges, which are modelled as piecewise constant. These discussions about the single-particle wavefunctions are actually quite standard\cite{LLsinHgTeQW,Wuerzburg-band-structure}. 

The other LL that is closest to the bulk gap is $\psi_0$, which has seven nonzero components, including three $n=0$ components, three $n=1$ components and one $n=2$ component. Because of the $n=1$ components it may have  some similar behaviours to the $n=1$ LL in conventional 2D electron gas, e.g. supporting the Moore-Read state\cite{Pf} and the Read-Rezayi state\cite{RRp1}.

If we ignore LL mixing, then the effect of the finite width and the mutli-component nature of the single-particle wavefunction can be absorbed into a form factor in analogy to the four-band case in Ref.~\onlinecite{ME}. Since we use a different geometry (torus instead of disk) and hence a different gauge (Landau instead of the symmetric gauge), it may be worthwhile to repeat some of the steps. The single-particle wavefunction is actually, after the magnetic translational invariance along the two edges of the system $\vL_1=(L_1,0)$ and $\vL_2=(L_{2x},L_{2y})$ (in the $xy$ plane)\cite{BR} is taken into account, 
\begin{widetext}
\begin{equation}
\phi_{\nLL, j_x}=C\sum_{k\in\ZZ}\left\{\left.\left[e^{-\frac{\ti}{2}\frac{\Ltx}{\Lty}\ell^2k_x^2}\psi_{\nLL,k_x}(\vcr)\right]\right\vert _{k_x=2\pi(j_x+k\Nf)/L_1}\right\}.
\end{equation}
In the above expression, $C$ is a normalization constant, $\psi_{\nLL,k_x}(\vcr)$ is as given in Eq.~(\ref{envelop-func}), $\Nf$ is the number of magnetic fluxes through the system, and $j_x=1,2,\dots,\Nf$ replaces $k_x$ as one of the single-particle quantum numbers. The matrix element of the Fourier transform of the density is given by 

\begin{equation}
u_{\nLL,j_{x2};\nLL,j_{x1}}(\vQ)=\int\td \vcr \phi_{\nLL,j_{x2}}^{\dagger}(\vcr)e^{\ti\vec{Q}\cdot \vcr}\phi_{\nLL,j_{x1}}(\vcr).\label{u-in-terms-of-integral}
\end{equation}
Note that $\vcr$ and $\vQ$ are three-dimensional vectors. In particular, $\vQ=\vQp+Q_z\hat{z}$, where $\vQp$ is a vector in the $xy$ plane and is limited to a lattice
\begin{align}
\vQp=&m\vec{q}_1+\tn \vec{q}_2,\:\: m,\tn\in \ZZ,\notag \\
\vq_1=&\frac{2\pi}{A}\vL_2\times \hat{z}=\frac{2\pi}{L_1\Lty}(\Lty \hat{x}-\Ltx \hat{y})=2\pi \left(\frac{\hat{x}}{L_1}-\frac{\Ltx}{L_1\Lty}\hat{y}\right),\label{q1}\\
\vq_2=&-\frac{2\pi}{A}\vL_1\times\hat{z}=\frac{2\pi}{L_1\Lty}L_1\hat{y}=\frac{2\pi}{\Lty}\hat{y}.\label{q2}
\end{align}
After performing the 2D integral in the $xy$ plane (over the parallelogram determined by $\vL_1$ and $\vL_2$), we get (suppressing the label $\nLL$)
\begin{equation}
u_{j_{x2},j_{x1}}(\vQ)=|C|^2\int^{+\infty}_{-\infty}\td z \sum_i |f_i(z)|^2 e^{\ti Q_z z}L_1 \delta'_{j_{x2},j_{x1}+m}\exp\left\{\ti \frac{\pi}{\Nf}(2j_{x2}-m)\tn-\frac{1}{4} \Qp^2\ell^2\right\}L_{n_i}\left(\half \Qp^2\ell^2\right),
\end{equation}
where the prime on the Kronecker delta means that the two subscripts are equal mod $\Nf$, $L_{n_i}\left(\half \Qp^2\ell^2\right)$ are the Laguerre polynomials, and explicitly $\Qp^2=4\pi^2\left[\frac{1}{\Lty^2}(-\frac{m\Ltx}{L_1}+\tn)^2+\frac{m^2}{L_1^2}\right]$. Setting $\vQ=0$ in the above expression gives us\[
\delta'_{j_{x2},j_{x1}}=|C|^2\int^{+\infty}_{-\infty}\td z\sum_i |f_i(z)|^2 L_1\delta'_{j_{x2},j_{x1}}.
\] 
Writing $C=\frac{1}{\sqrt{L_1}}\bC$,  we find that the normalization constant $\bC$ is determined by \begin{equation}
1=|\bC|^2\int^{+\infty}_{-\infty}\td z \sum_i |f_i(z)|^2 \label{f-i-normalization}\;.
\end{equation}
Defining\begin{equation}
F(\Qp,z)=|\bC|^2\sum_i |f_i(z)|^2 L_{n_i}(\half \Qp^2\ell^2)\label{F-definition},
\end{equation}
then the $u$ matrix element can be written \begin{equation}
u_{j_{x2},j_{x1}}=\delta'_{j_{x2},j_{x1}+m}\left[\int^{+\infty}_{-\infty}\td z F(\Qp,z) e^{\ti Q_z z}\right] \exp\left\{\ti \frac{\pi}{\Nf}(2j_{x2}-m)\tn-\frac{1}{4} \Qp^2\ell^2\right\}.\label{u-finite-width-in-terms-of-F}
\end{equation}
The Coulomb matrix elements are, shortening the notations $j_{x1}$ etc. to $j_1$ etc., \begin{equation}
V_{j_1,j_2,j_3,j_4}=\frac{1}{A}\sum'_{m,\tn\in\ZZ}\int^{+\infty}_{-\infty}\frac{\td Q_z}{2\pi} \frac{4\pi e^2}{\epsilon Q^2}u_{j_1,j_4}(-\vQ)u_{j_2,j_3}(\vQ),\label{V-3D-in-terms-of-u}
\end{equation}
where $A$ is the system size (area in the $xy$ plane), the prime on the summation sum indicates that the term $m=\tn=0$ should be omitted\cite{Yoshioka1,Su}, and $\epsilon$ is the dialectric constant. Substituting Eq.~(\ref{u-finite-width-in-terms-of-F}) in Eq.~(\ref{V-3D-in-terms-of-u}), we get\begin{align}
V_{j_1,j_2,j_3,j_4}=&\frac{1}{A}\sum'_{m,\tn\in\ZZ}\int^{+\infty}_{-\infty}\frac{\td Q_z}{2\pi} \frac{4\pi e^2}{\epsilon Q^2}\int^{+\infty}_{-\infty}\td z_1  F(\Qp,z_1) e^{-\ti Q_z z_1}\delta'_{j_1,j_4-m}e^{\ti\frac{\pi}{\Nf}(2j_1+m)(-\tn)-\frac{1}{4}\Qp^2\ell^2}\label{V-finite-width-expression-1}\\
&\int^{+\infty}_{-\infty}\td z_2 F(\Qp,z_2) e^{\ti Q_z z_2}\delta'_{j_2,j_3+m}e^{\ti\frac{\pi}{\Nf}(2j_2-m)\tn-\frac{1}{4}\Qp^2\ell^2} .\notag
\end{align}
The $Q_z$ integral can be performed easily,\[
\int^{+\infty}_{-\infty}\frac{\td Q_z}{2\pi} \frac{e^{\ti Q_z (z_2-z_1)}}{Q_z^2+\Qp^2}=\frac{e^{-  \Qp|z_1-z_2|}}{2\Qp}.
\]
Then
\begin{equation}
V_{j_1,j_2,j_3,j_4}=\delta'_{j_1+j_2,j_3+j_4}\frac{1}{A}\sum'_{k,\tn\in\ZZ}\left\{\left.\left[ \frac{2\pi e^2}{\epsilon \Qp}\scF(\Qp,n_{\text{LL}})e^{\ti \frac{2\pi}{\Nf}(j_2-j_1-m)\tn-\half \Qp\ell^2}\right]\right\vert_{m=j_2-j_3+k\Nf}\right\},\label{V-finite-width-expression2}
\end{equation}
where the $\sum'$ now means that $\tn$ and $j_2-j_3+k\Nf$ should not be simultaneously zero, and we have defined \begin{equation}
\scF(\Qp,n_{\text{LL}})=\int^{+\infty}_{-\infty}\td z_1 \td z_2 e^{-\Qp |z_1-z_2|}F(\Qp, z_1)F(\Qp,z_2)\label{def-scF-no-LLM}.
\end{equation}
The effect of the complicated band structure of the single-particle problem has been absorbed into this form factor, as advertised. The interaction part of the Hamiltonian are given in terms of the $V$ elements as $H_{\text{int}}=\half\sum_{j_1,j_2,j_3,j_4}V_{j_1,j_2,j_3,j_4}\cdg_{j_1}\cdg_{j_2}c_{j_3}c_{j_4}$, where the $\cdg$ and $c$ are creation and annihilation operators.

In the vicinity of the level crossing point, we will take into account LL mixing by doing exact diagonalization with both LLs included. In this case we will need inter-LL matrix elements and a straightforward generalization of Ref.~\onlinecite{ME} is carried out. Specifically, the inter-LL $u$ matrix elements are \begin{align}
u_{\nLLb,j_{x2};\nLLa,j_{x1}}(\vQ)=&\int\td \vcr \phi_{\nLLb,j_{x2}}^{\dagger}(\vcr)e^{\ti\vec{Q}\cdot \vcr}\phi_{\nLLa,j_{x1}}(\vcr)\label{u-in-terms-of-integral-interLL}\\
=&\delta'_{j_{x2},j_{x1}+m}\left[\int^{+\infty}_{-\infty}\td z F^{(\nLLb,\nLLa)}(\vQp,z)e^{\ti Q_z z}\right]\exp\left\{\ti \frac{\pi}{\Nf}(2j_{x2}-m)\tn-\frac{1}{4}\Qp^2\ell^2\right\}\label{u3D-LLM-in-terms-of-F},
\end{align}
where we have defined \begin{align}
&F^{(\nLLb,\nLLa)}(\vQp,z)\notag \\
=&\bC_{\nLLb}^*\bC_{\nLLa}\sum_{i=1}^8 f^{(\nLLb)*}_i(z)f^{(\nLLa)}_i(z)\frac{(-1)^{(n_i^{(\nLLb)}+n_i^{(\nLLa)})}2^{\max(n_i^{(\nLLb)},n_i^{(\nLLa)})}  [\min(n_i^{(\nLLb)},n_i^{(\nLLa)})]!}{\left[2^{(n_i^{(\nLLb)}+n_i^{(\nLLa)})}n_i^{(\nLLb)}!n_i^{(\nLLa)}!\right]^{1/2}} \label{F-finite-width-LLM}\\
&\times\left\{ \pi\left[\frac{\ti\ell}{\Lty}\left(-\frac{m L_{2x}}{L_1}+\tn\right)+\sgn (n_i^{(\nLLb)}-n_i^{(\nLLa)})\frac{m\ell}{L_1} \right] \right\}^{| n_i^{(\nLLb)}-n_i^{(\nLLa)} |}L^{| n_i^{(\nLLb)}-n_i^{(\nLLa)} |}_{\min(n_i^{(\nLLb)},n_i^{(\nLLa)})}\left(\half\Qp^2\ell^2\right),\notag
\end{align}
with the last factor being the associated Laguerre polynomial. The form factor for $V_{\nLLa,j_1;\nLLb,j_2;\nLLc,j_3;\nLLd,j_4}$ is \begin{equation}
\scF^{(\nLLa,\nLLb,\nLLc,\nLLd)}(\vQp)=\int ^{+\infty}_{-\infty}\td z_1\td z_2 e^{-\Qp |z_1-z_2|}F^{(\nLLa,\nLLd)}(-\vQp,z_1)F^{(\nLLb,\nLLc)}(\vQp,z_2)\label{def-scF-LLM}
\end{equation}
[this replaces the $\scF(\Qp,n_{\text{LL}})$ in Eq.~(\ref{V-finite-width-expression2}), with the other factors unchanged]. Now the interaction part of the Hamiltonian is $H_{\text{int}}=\half\sum_{\nLLa,j_1;\nLLb,j_2;\nLLc,j_3;\nLLd,j_4}V_{\nLLa,j_1;\nLLb,j_2;\nLLc,j_3;\nLLd,j_4}\cdg_{\nLLa,j_1}\cdg_{\nLLb,j_2}c_{\nLLc,j_3}c_{\nLLd,j_4}$.
\end{widetext}

\begin{figure}[t!]
\begin{center}
\includegraphics[scale=.8]{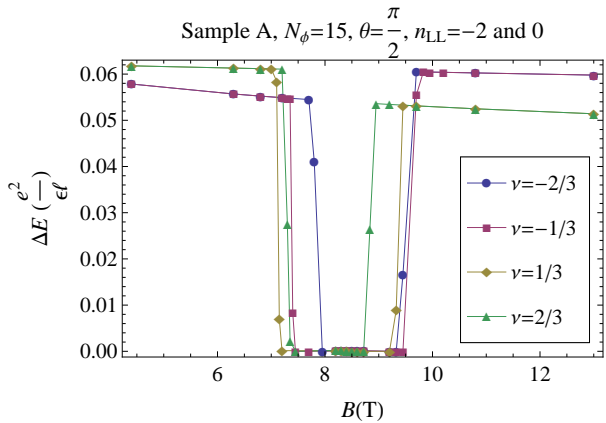}
\end{center}
\caption{\label{fig:gap-vs-B-13} (color online) Gaps as functions of magnetic field for $\nu=\pm 1/3, \pm 2/3$ in sample A as calculated by exact diagonalization. Both $\nLL=-2$ and $\nLL=0$ levels are included in the diagonalization. The magnetic flux through the unit cell is $\Nf=15$ flux quanta (by unit cell we mean the simulation cell, i.e. the parallelogram with $\vL_1$ and $\vL_2$ as two adjacent sides, not the unit cell of the atomic lattice). The aspect ratio, i.e. $|\vL_2/\vL_1|=1$ is chosen to be 1, and the angel between $\vL_1$ and $\vL_2$ (denoted $\theta$) is chosen to be $\pi/2$, i.e. the unit cell is a square. 
}
\end{figure}

\begin{figure}
\begin{center}
\subfigure[]{\label{fig:tnu13nLLm2}\includegraphics[scale=.8]{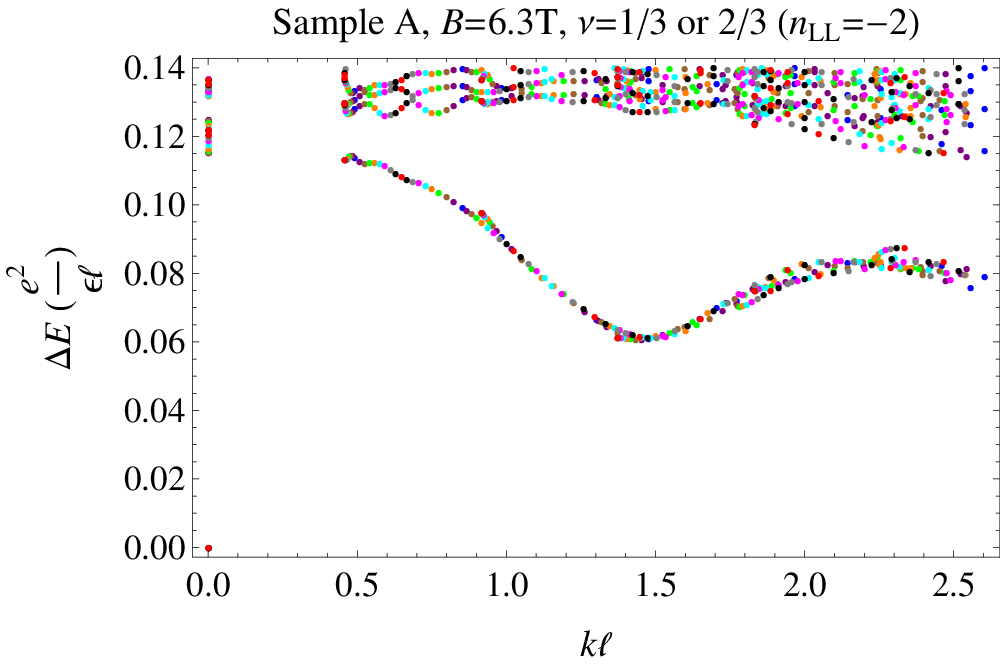}}
\subfigure[]{\label{fig:tnu13nLL0}\includegraphics[scale=.8]{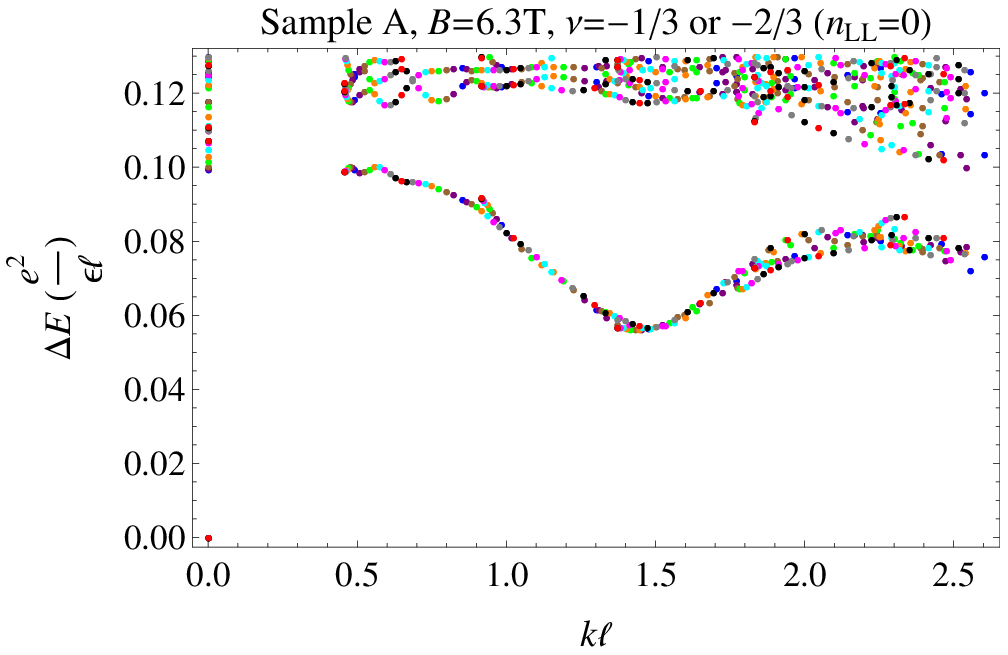}}
\end{center}
\caption{\label{fig:13like} (color online) Exact diagonalization spectra for sample A of Ref.~\onlinecite{LLsinHgTeQW} at $B=6.3$T for (a) $\nu=1/3$ and $2/3$ (the fractionally filled LL is $\nLL=-2$), and (b) $\nu=-1/3$ and $-2/3$ (the fractionally filled LL is $\nLL=0$). For all the data in this figure we used $\Nf=30$ and $|\vL_1|/|\vL_2|=1$, and each color corresponds to a different angle $\theta\in[\frac{\pi}{3},\frac{\pi}{2}]$. For the horizontal axis, $k$ is the magnitude of $\vk$, which is a wavevectorlike many-body quantum number characterizing the relative motions of the electrons\cite{Haldane,BR} (and has no relation with the $k_x$ earlier in the discussion about the single-particle wavefunction).
}
\end{figure}

\begin{figure}
\begin{center}
\includegraphics[scale=.8]{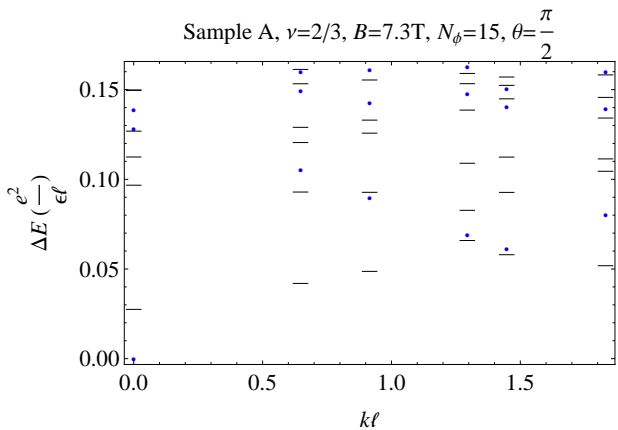}
\end{center}
\caption{\label{fig:nu23spectrum} (color online) The spectrum for sample A with $\nu=2/3$, $B=7.3$T, and $\Nf=15$. Both $\nLL=-2$ and $\nLL=0$ LLs are included in the diagonalization (the total number of eletrons in these two LLs is then 25). The dots are levels with $\langle N_{-2}\rangle\approx 10$ electrons in the $\nLL=-2$ LL (the higher level at this $B$). The dashes are levels with $\langle N_{-2}\rangle \approx 11$ electrons in the $\nLL=-2$ LL (i.e. one electron is promoted to this LL from the $\nLL=0$ LL, leaving a hole in the latter). 
}
\end{figure}

\begin{figure}
\begin{center}
\subfigure[]{\label{fig:nu12Ne10}\includegraphics[scale=.8]{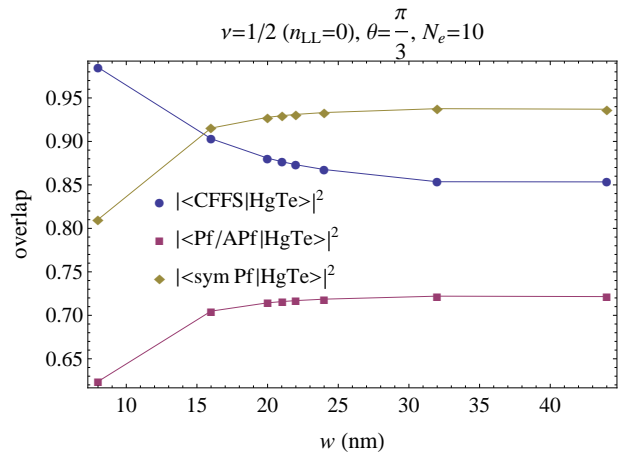}}
\subfigure[]{\label{fig:n12Ne12}\includegraphics[scale=.8]{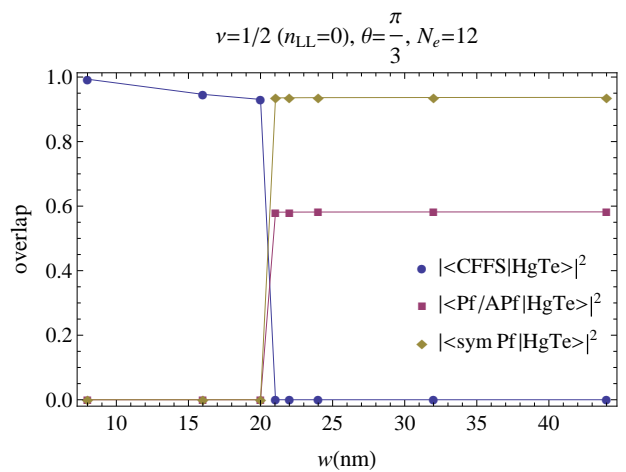}}
\subfigure[]{\label{fig:nu12Ne14overlaps-v2}\includegraphics[scale=.8]{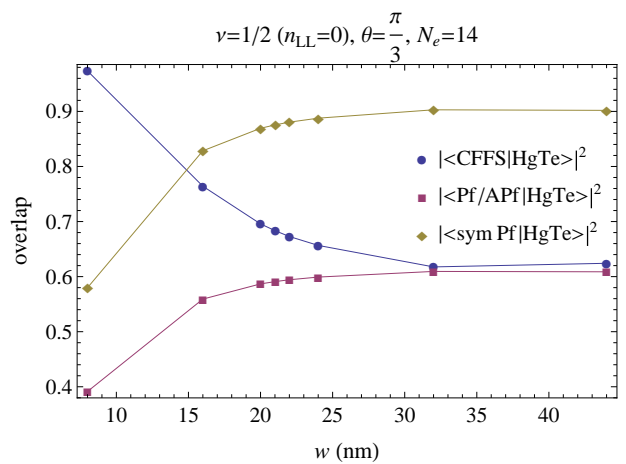}}
\end{center}
\caption{\label{fig:nu12overlap4s} (color online) Overlaps between the Coulomb GS in HgTe quantum well with the Pf/APf, the symPf and the CFFS for $\nu=1/2$ and (a) $\Ne=10$, (b) $\Ne=12$, and (c) $\Ne=14$, as functions of the well width $w$.  
}
\end{figure}

We first consider sample A in Ref.~\onlinecite{LLsinHgTeQW} and $1/3$-like states. For sample A, we do not know the electron density (which is due to a small residual doping), so we consider many different possibilities, which is equivalent to treating the density as a tunable parameter. Since the $\nLL=-2$ and $\nLL=0$ LLs cross each other at $B_{\text{c}}\simeq 8.72$T, LL mixing may be important. Therefore we include both LLs simultaneously in the calculation \cite{Yoshioka}. Fig.~\ref{fig:gap-vs-B-13} shows the gap as a function of the magnetic field for $\nu=\pm 1/3, \pm 2/3$ ($\nu=0$ corresponds to, at least in the noninteracting case, the lower of these two LLs completely filled, and the higher one completely empty). For $B$ sufficiently away from $B_{\text{c}}$ on each side we find that the effect of LL mixing vanishes. This is indicated by the fact that $\nu=1/3$ and $\nu=2/3$ have the same spectrum [i.e. particle-hole symmetry within the $\nLL=-2$ (0) level for $B<(>)\Bc$], which is also the same as that from a single-LL calculation. Similarly, $\nu=-1/3$ and $\nu=-2/3$ have the same spectrum [particle-hole symmetry within the $\nLL=0$ ($-2$) level for $B<(>)\Bc$]. In these regions we can safely use a single LL for the calculation and go to larger system size (i.e. larger $\Nf$). The spectra for $B=6.3$T are shown in Fig.~\ref{fig:13like} as an example. The neutral gaps are determined by the minima of the magnetoroton branches (the low-lying curve in each subfigure). We note in passing that the dispersion of the magnetoroton branch can be measured experimentally \cite{Kukushkin22052009}. When $B$ is closer to $\Bc$, for each filling factor there is a narrow region (on each side of $\Bc$) where the gap is determined by inter-LL excitations (excitations where some electrons are excited to the higher LL from the otherwise filled lower LL) and is reduced compared to the magnetoroton minimum, see e.g. Fig.~\ref{fig:nu23spectrum}. For $B$ even closer to $\Bc$ the gap is totally destroyed by LL mixing.

\begin{figure}
\begin{center}
\includegraphics[scale=.8]{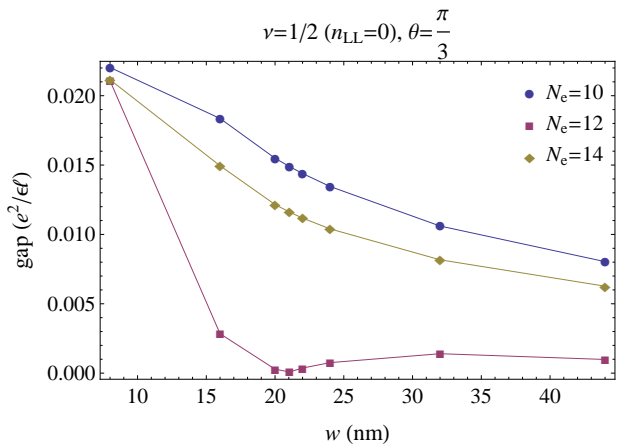}
\end{center}
\caption{\label{fig:nu12gaps} (color online) Gaps for $\nu=1/2$ as a function of well width for $\Ne=10,12,14$. 
}
\end{figure}

\begin{figure}
\begin{center}
\includegraphics[scale=.8]{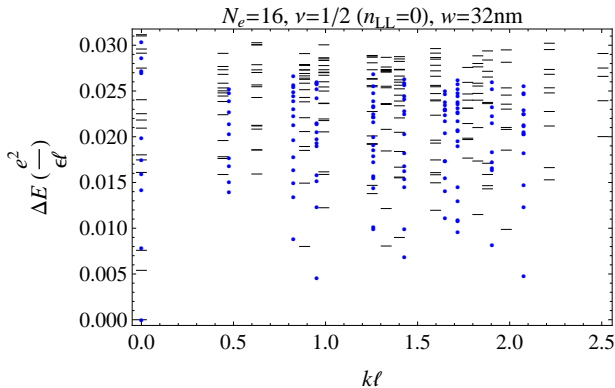}
\end{center}
\caption{\label{fig:nu12spectrum} (color online) Spectrum for $\nu=1/2$, $w=32$nm, with $\Ne=16$. The blue dots are for the hexagonal unit cell, while the black dashes are for the square unit cell. The scheme of Ref.~\onlinecite{Papic-Haldane} is used to fold the Brillouin zone to a quarter of the original size to account for the pairing nature of the Pf/APf state. In particular the GSs are mapped from the centers of the edges in the original Brillouin zone to the center of the folded one ($\vk=0$). For the hexagon unit cell, the GS's overlaps with the Pf/APf and the symPf are 0.50 and 0.86, respectively [the CFFS for this system size and geometry has $\vk=-4\vq_1-4\vq_2$ (and $\vk$'s related by rotational symmetry) in the unfolded Brillouin zone, which is different from the Coulomb state here, so there is no overlap]. For the square unit cell, the overlaps between the GS and the CFFS, the Pf/APf, and the symPf are 0.50, 0.55 and 0.86, respectively.
}
\end{figure}

Next we consider $\nu=1/2$. To be concrete, we fix the electron density to $4.2\times 10^{11}\text{cm}^{-2}$ (as for sample B in Ref.~\onlinecite{LLsinHgTeQW}). We consider several different well widths $w$. For the density we choose, $\nu=1/2$ corresponds to $B=34.75$T and is to the right of the level crossing point between $\nLL=-2$ and $\nLL=0$ for all the well widths and therefore the fractionally filled LL is the $\nLL=0$ level (and for all well widths this level is sufficiently far away from other levels for LL mixing to vanish, so we will only keep this one level in the exact diagonalization). 

First we look at $w=8$nm as for sample B of Ref.~\onlinecite{LLsinHgTeQW}. Unless otherwise specified the geometry of the unit cell is hexagonal. For $\Ne=10$ , the ground state (GS) has $\vk=5\vq_1$ [$\vq_1$ and $\vq_2$ are the basis vectors defined in Eqs.~(\ref{q1}) and (\ref{q2})] and the two $\vk$'s related by rotational symmetry, i.e. $5\vq_2$ and $5(\vq_1+\vq_2)$ [i.e. the GS is three-fold degenerate (not counting the trivial two-fold center-of-mass degeneracy\cite{Haldane,RRp0})]. These three $\vk$'s correspond to the centers of the edges of the Brillouin zone. This behavior is consistent\cite{RRp0} with the Moore-Read Pfaffian (Pf) state\cite{Pf} [or the anti-Pfaffian\cite{aPf-Levin,aPf-Lee} (APf), the particle-hole conjugate of the Pf]. The overlap with the Pfaffian state is considerable at 0.62 (the overlap with the APf is the same in the absence of LL mixing). In Ref.~\onlinecite{Rezayi-Haldane} it was argued that since the Coulomb state is particle-hole symmetric in the absence of LL mixing, while the Pf state is not, we should consider the particle-hole symmetrized Pf (symPf) instead. In the current case, the overlap with the symPf is 0.81. However, this does not necessarily mean that the Coulomb state is the symPf. The other possibility is the composite fermion Fermi sea (CFFS)\cite{Rezayi-Read-state}, also known as the Rezayi-Read state (not to be confused with the Read-Rezayi state to be discussed below). There is no known parent Hamiltonian, i.e. a Hamiltonian for which this state is exactly the GS, so to calculate the overlap with this state one would need to use Monte Carlo method, which would take too long for the system sizes we are using here. Fortunately, it is known (from Monte Carlo calculations with smaller system sizes) that the Coulomb state in the lowest LL in the purely 2D case [i.e. $H_{\text{para,mag}}$ given below Eq.~(\ref{envelop-func})] has overlap with the CFFS that is practically 1\cite{Rezayi-Read-state,Rezayi-Haldane}, so assuming this is true also for larger system size, one can approximate the CFFS with the lowest LL Coulomb state\cite{Morf,LLLCoulombState-as-CFFS-p2} (the latter has the added advantage of being exactly particle-hole symmetric, while for the former the particle-hole symmetry is only approximate\cite{Rezayi-Haldane}). With this approximation, we find that in the current case the overlap with the CFFS is 0.99, much higher than that with the symPf. For $\Ne=12$ the GS has $\vk=4\vq_1+2\vq_2$ (and some other values related by rotational symmetry) and is not consistent with the Pf/APf/symPf. The overlap with the CFFS is 0.99. 

\begin{figure}
\begin{center}
\includegraphics[scale=.8]{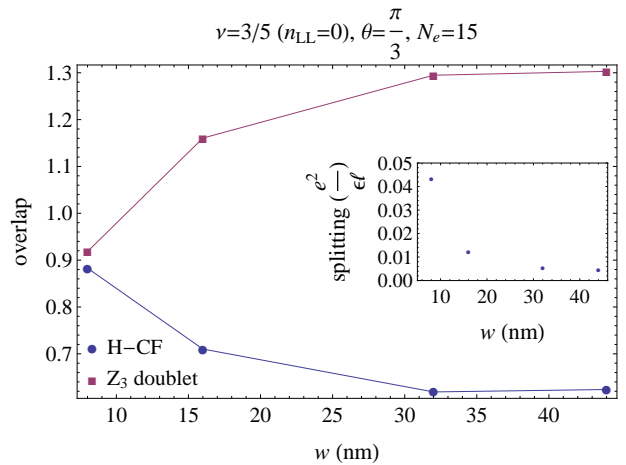}
\end{center}
\caption{\label{fig:nu35overlap-splitting} (color online) The overlap between the GS and the H-CF state and the total overlap of the GS and lowest excited state (at $\vk=0$) with the $Z_3$ doublet for $\nu=3/5$ as functions of the well width. The inset is the splitting between the GS and the lowest excited state (at $\vk=0$).
}
\end{figure}

\begin{figure}
\begin{center}
\includegraphics[scale=.8]{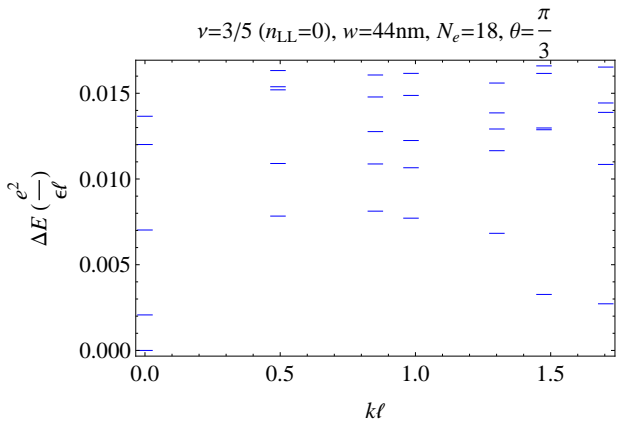}
\end{center}
\caption{\label{fig:nu35spectrum} (color online) Exact diagonalization spectrum for $\nu=3/5$ and $w=44$nm with $\Ne=18$
}
\end{figure}

For $\Ne=14$ the GS has $\vk$ consistent with the Pf/APf/symPf, the overlaps with the Pf/APf, the symPf and the CFFS are 0.39, 0.58 and 0.97, respectively. The results for all three system sizes suggest that at this value of well width ($w=8$nm) the GS is the CFFS. However, when we increase the well width, for $\Ne=10$ and 14, the overlaps with the Pf and the symPf increases while the overlap with the CFFS decreases, and for $w>w_c\approx 15$nm, the overlap with the symPf is larger than those with the CFFS, see Figs.~\ref{fig:nu12Ne10} and \ref{fig:nu12Ne14overlaps-v2}. 

For $\Ne=12$, the $\vk$ of the GS changes from that corresponding to the CFFS to that corresponding to the Pf/APf/symPf at around $w=21$nm [Fig.~\ref{fig:n12Ne12}]. These behaviours suggest a crossover from the CFFS at small well width to the symPf at large well width.  In Fig.~\ref{fig:nu12gaps} we show the well width dependence (and system size dependence) of the gap. In Fig.~\ref{fig:nu12spectrum} we show the spectrum for $\Ne=16$ (both hexagon and square unit cells) at $w=32$nm. The GS $\vk$ values and the overlaps for this system size also suggest the symPf state.

Finally, we consider $\nu=3/5$. This filling factor is of particular interest because the GS could be in the universality class of the $Z_3$ parafermion state \cite{RRp1}, which can support universal quantum computation \cite{uqc}. An competing state for this filling factor is the hierarchy state\cite{Haldane-Hierarchy,Halperin-Hierarchy}, which can also be described as an integer quantum Hall effect of composite fermions\cite{Jain}. As for $\nu=1/2$, we fix the density to $4.2\times 10^{11}\text{cm}^{-2}$ and vary the well width. For this density $\nu=3/5$ corresponds to $B=28.96$T, for which the $\nLL=0$ LL is also far away from other LLs, so we only include this one LL in the exact diagonalization. For $w=8$nm, the overlap between the GS and the hierarchy-CF (H-CF) state is 0.88. The GS for the $Z_3$ state is two-fold degenerate (the GS $\vk$ is equal to 0). For HgTe quantum well the GS is nondegenerate (not including the trivial five-fold center-of-mass degeneracy) and the splitting between the Coulomb GS and the lowest excited state (at $\vk=0$) is $0.043\ec$, which is quite large. However, if we ignore this splitting and calculate the total overlap of the GS and the first excited state with the $Z_3$ doublet\cite{RRp1}, i.e. $\sum_{i=1}^2(|\langle Z_3,i|\text{GS}\rangle|^2+|\langle Z_3,i|\text{FES}\rangle|^2)$, where $|Z_3,1\rangle$ and $|Z_3,2\rangle$ are orthonormal basis states of the $Z_3$ GS subspace, $|\text{GS}\rangle$ and $|\text{FES}\rangle$ denote the GS and the first excited state (at $\vk=0$) for the HgTe quantum well, we get 0.92. Hence for $w=8$nm neither the H-CF state nor the $Z_3$ state is clearly favoured. As $w$ is increased, the overlap between the GS and the H-CF state decreases, the splitting between the GS and the first excited state decreases, and the total overlap of these two states with the $Z_3$ GS doublet increases, see Fig.~\ref{fig:nu35overlap-splitting}. Therefore, at least for the larger well widths, the $Z_3$ state is a viable candidate for the GS at $\nu=3/5$. In Fig.~\ref{fig:nu35spectrum} we show the spectrum for $w=44$nm and $\Ne=18$.

To conclude, our exact diagonalization results show that HgTe quantum wells are capable of supporting fractional quantum Hall effects. The filling factors we studied include $\nu=\pm 1/3, \pm 2/3, 1/2$, and $3/5$. For $\nu=1/2$ we find a crossover from the composite fermion Fermi sea at small well width to the particle-hole symmetrized Moore-Read state. For $\nu=3/5$, the Read-Rezayi $Z_3$ state dominates over the hierarchy state at large well width but gradually loses this dominance when the well width is decreased. We also find the effect of Landau level mixing can be important near the level crossing point and can totally destroy the gap. We note that our study does not take into account disorder, which can reduce the gaps significantly and make the experimental observation difficult.

{\it Acknowledgement}--I thank Christoph Br\"une for a useful discussion. This work was partly supported by the Kreitman Foundation.


\begin{thebibliography}{26}%
\makeatletter
\providecommand \@ifxundefined [1]{%
 \@ifx{#1\undefined}
}%
\providecommand \@ifnum [1]{%
 \ifnum #1\expandafter \@firstoftwo
 \else \expandafter \@secondoftwo
 \fi
}%
\providecommand \@ifx [1]{%
 \ifx #1\expandafter \@firstoftwo
 \else \expandafter \@secondoftwo
 \fi
}%
\providecommand \natexlab [1]{#1}%
\providecommand \enquote  [1]{``#1''}%
\providecommand \bibnamefont  [1]{#1}%
\providecommand \bibfnamefont [1]{#1}%
\providecommand \citenamefont [1]{#1}%
\providecommand \href@noop [0]{\@secondoftwo}%
\providecommand \href [0]{\begingroup \@sanitize@url \@href}%
\providecommand \@href[1]{\@@startlink{#1}\@@href}%
\providecommand \@@href[1]{\endgroup#1\@@endlink}%
\providecommand \@sanitize@url [0]{\catcode `\\12\catcode `\$12\catcode
  `\&12\catcode `\#12\catcode `\^12\catcode `\_12\catcode `\%12\relax}%
\providecommand \@@startlink[1]{}%
\providecommand \@@endlink[0]{}%
\providecommand \url  [0]{\begingroup\@sanitize@url \@url }%
\providecommand \@url [1]{\endgroup\@href {#1}{\urlprefix }}%
\providecommand \urlprefix  [0]{URL }%
\providecommand \Eprint [0]{\href }%
\providecommand \doibase [0]{http://dx.doi.org/}%
\providecommand \selectlanguage [0]{\@gobble}%
\providecommand \bibinfo  [0]{\@secondoftwo}%
\providecommand \bibfield  [0]{\@secondoftwo}%
\providecommand \translation [1]{[#1]}%
\providecommand \BibitemOpen [0]{}%
\providecommand \bibitemStop [0]{}%
\providecommand \bibitemNoStop [0]{.\EOS\space}%
\providecommand \EOS [0]{\spacefactor3000\relax}%
\providecommand \BibitemShut  [1]{\csname bibitem#1\endcsname}%
\let\auto@bib@innerbib\@empty
\bibitem [{\citenamefont {Bernevig}\ \emph {et~al.}(2006)\citenamefont
  {Bernevig}, \citenamefont {Hughes},\ and\ \citenamefont {Zhang}}]{BHZ}%
  \BibitemOpen
  \bibfield  {author} {\bibinfo {author} {\bibfnamefont {B.~A.}\ \bibnamefont
  {Bernevig}}, \bibinfo {author} {\bibfnamefont {T.~L.}\ \bibnamefont
  {Hughes}}, \ and\ \bibinfo {author} {\bibfnamefont {S.-C.}\ \bibnamefont
  {Zhang}},\ }\href {\doibase 10.1126/science.1133734} {\bibfield  {journal}
  {\bibinfo  {journal} {Science}\ }\textbf {\bibinfo {volume} {314}},\ \bibinfo
  {pages} {1757} (\bibinfo {year} {2006})}\BibitemShut {NoStop}%
\bibitem [{\citenamefont {K\"onig}\ \emph {et~al.}(2007)\citenamefont
  {K\"onig}, \citenamefont {Wiedmann}, \citenamefont {Br\"une}, \citenamefont
  {Roth}, \citenamefont {Buhmann}, \citenamefont {Molenkamp}, \citenamefont
  {Qi},\ and\ \citenamefont {Zhang}}]{Molenkamp}%
  \BibitemOpen
  \bibfield  {author} {\bibinfo {author} {\bibfnamefont {M.}~\bibnamefont
  {K\"onig}}, \bibinfo {author} {\bibfnamefont {S.}~\bibnamefont {Wiedmann}},
  \bibinfo {author} {\bibfnamefont {C.}~\bibnamefont {Br\"une}}, \bibinfo
  {author} {\bibfnamefont {A.}~\bibnamefont {Roth}}, \bibinfo {author}
  {\bibfnamefont {H.}~\bibnamefont {Buhmann}}, \bibinfo {author} {\bibfnamefont
  {L.~W.}\ \bibnamefont {Molenkamp}}, \bibinfo {author} {\bibfnamefont {X.-L.}\
  \bibnamefont {Qi}}, \ and\ \bibinfo {author} {\bibfnamefont {S.-C.}\
  \bibnamefont {Zhang}},\ }\href@noop {} {\bibfield  {journal} {\bibinfo
  {journal} {Science}\ }\textbf {\bibinfo {volume} {318}},\ \bibinfo {pages}
  {766} (\bibinfo {year} {2007})}\BibitemShut {NoStop}%
\bibitem [{\citenamefont {Yoshioka}(1984)}]{Yoshioka1}%
  \BibitemOpen
  \bibfield  {author} {\bibinfo {author} {\bibfnamefont {D.}~\bibnamefont
  {Yoshioka}},\ }\href {\doibase 10.1103/PhysRevB.29.6833} {\bibfield
  {journal} {\bibinfo  {journal} {Phys. Rev. B}\ }\textbf {\bibinfo {volume}
  {29}},\ \bibinfo {pages} {6833} (\bibinfo {year} {1984})}\BibitemShut
  {NoStop}%
\bibitem [{\citenamefont {Su}(1984)}]{Su}%
  \BibitemOpen
  \bibfield  {author} {\bibinfo {author} {\bibfnamefont {W.~P.}\ \bibnamefont
  {Su}},\ }\href {\doibase 10.1103/PhysRevB.30.1069} {\bibfield  {journal}
  {\bibinfo  {journal} {Phys. Rev. B}\ }\textbf {\bibinfo {volume} {30}},\
  \bibinfo {pages} {1069} (\bibinfo {year} {1984})}\BibitemShut {NoStop}%
\bibitem [{\citenamefont {Haldane}(1985)}]{Haldane}%
  \BibitemOpen
  \bibfield  {author} {\bibinfo {author} {\bibfnamefont {F.~D.~M.}\
  \bibnamefont {Haldane}},\ }\href {\doibase 10.1103/PhysRevLett.55.2095}
  {\bibfield  {journal} {\bibinfo  {journal} {Phys. Rev. Lett.}\ }\textbf
  {\bibinfo {volume} {55}},\ \bibinfo {pages} {2095} (\bibinfo {year}
  {1985})}\BibitemShut {NoStop}%
\bibitem [{\citenamefont {Bernevig}\ and\ \citenamefont {Regnault}(2012)}]{BR}%
  \BibitemOpen
  \bibfield  {author} {\bibinfo {author} {\bibfnamefont {B.~A.}\ \bibnamefont
  {Bernevig}}\ and\ \bibinfo {author} {\bibfnamefont {N.}~\bibnamefont
  {Regnault}},\ }\href {\doibase 10.1103/PhysRevB.85.075128} {\bibfield
  {journal} {\bibinfo  {journal} {Phys. Rev. B}\ }\textbf {\bibinfo {volume}
  {85}},\ \bibinfo {pages} {075128} (\bibinfo {year} {2012})}\BibitemShut
  {NoStop}%
\bibitem [{\citenamefont {Novik}\ \emph {et~al.}(2005)\citenamefont {Novik},
  \citenamefont {Pfeuffer-Jeschke}, \citenamefont {Jungwirth}, \citenamefont
  {Latussek}, \citenamefont {Becker}, \citenamefont {Landwehr}, \citenamefont
  {Buhmann},\ and\ \citenamefont {Molenkamp}}]{Wuerzburg-band-structure}%
  \BibitemOpen
  \bibfield  {author} {\bibinfo {author} {\bibfnamefont {E.~G.}\ \bibnamefont
  {Novik}}, \bibinfo {author} {\bibfnamefont {A.}~\bibnamefont
  {Pfeuffer-Jeschke}}, \bibinfo {author} {\bibfnamefont {T.}~\bibnamefont
  {Jungwirth}}, \bibinfo {author} {\bibfnamefont {V.}~\bibnamefont {Latussek}},
  \bibinfo {author} {\bibfnamefont {C.~R.}\ \bibnamefont {Becker}}, \bibinfo
  {author} {\bibfnamefont {G.}~\bibnamefont {Landwehr}}, \bibinfo {author}
  {\bibfnamefont {H.}~\bibnamefont {Buhmann}}, \ and\ \bibinfo {author}
  {\bibfnamefont {L.~W.}\ \bibnamefont {Molenkamp}},\ }\href {\doibase
  10.1103/PhysRevB.72.035321} {\bibfield  {journal} {\bibinfo  {journal} {Phys.
  Rev. B}\ }\textbf {\bibinfo {volume} {72}},\ \bibinfo {pages} {035321}
  (\bibinfo {year} {2005})}\BibitemShut {NoStop}%
\bibitem [{Jai()}]{Jain-book}%
  \BibitemOpen
  \href@noop {} {}\bibinfo {note} {J.~K. Jain, {\it Composite fermions}
  (Cambridge University Press, 2007), \S\S 3.1 and 3.2}\BibitemShut {NoStop}%
\bibitem [{\citenamefont {Orlita}\ \emph {et~al.}(2011)\citenamefont {Orlita},
  \citenamefont {Masztalerz}, \citenamefont {Faugeras}, \citenamefont
  {Potemski}, \citenamefont {Novik}, \citenamefont {Br\"une}, \citenamefont
  {Buhmann},\ and\ \citenamefont {Molenkamp}}]{LLsinHgTeQW}%
  \BibitemOpen
  \bibfield  {author} {\bibinfo {author} {\bibfnamefont {M.}~\bibnamefont
  {Orlita}}, \bibinfo {author} {\bibfnamefont {K.}~\bibnamefont {Masztalerz}},
  \bibinfo {author} {\bibfnamefont {C.}~\bibnamefont {Faugeras}}, \bibinfo
  {author} {\bibfnamefont {M.}~\bibnamefont {Potemski}}, \bibinfo {author}
  {\bibfnamefont {E.~G.}\ \bibnamefont {Novik}}, \bibinfo {author}
  {\bibfnamefont {C.}~\bibnamefont {Br\"une}}, \bibinfo {author} {\bibfnamefont
  {H.}~\bibnamefont {Buhmann}}, \ and\ \bibinfo {author} {\bibfnamefont
  {L.~W.}\ \bibnamefont {Molenkamp}},\ }\href {\doibase
  10.1103/PhysRevB.83.115307} {\bibfield  {journal} {\bibinfo  {journal} {Phys.
  Rev. B}\ }\textbf {\bibinfo {volume} {83}},\ \bibinfo {pages} {115307}
  (\bibinfo {year} {2011})}\BibitemShut {NoStop}%
\bibitem [{\citenamefont {Moore}\ and\ \citenamefont {Read}(1991)}]{Pf}%
  \BibitemOpen
  \bibfield  {author} {\bibinfo {author} {\bibfnamefont {G.}~\bibnamefont
  {Moore}}\ and\ \bibinfo {author} {\bibfnamefont {N.}~\bibnamefont {Read}},\
  }\href {\doibase http://dx.doi.org/10.1016/0550-3213(91)90407-O} {\bibfield
  {journal} {\bibinfo  {journal} {Nucl. Phys. B}\ }\textbf {\bibinfo {volume}
  {360}},\ \bibinfo {pages} {362 } (\bibinfo {year} {1991})}\BibitemShut
  {NoStop}%
\bibitem [{\citenamefont {Read}\ and\ \citenamefont {Rezayi}(1999)}]{RRp1}%
  \BibitemOpen
  \bibfield  {author} {\bibinfo {author} {\bibfnamefont {N.}~\bibnamefont
  {Read}}\ and\ \bibinfo {author} {\bibfnamefont {E.}~\bibnamefont {Rezayi}},\
  }\href {\doibase 10.1103/PhysRevB.59.8084} {\bibfield  {journal} {\bibinfo
  {journal} {Phys. Rev. B}\ }\textbf {\bibinfo {volume} {59}},\ \bibinfo
  {pages} {8084} (\bibinfo {year} {1999})}\BibitemShut {NoStop}%
\bibitem [{\citenamefont {MacDonald}\ and\ \citenamefont
  {Ekenberg}(1989)}]{ME}%
  \BibitemOpen
  \bibfield  {author} {\bibinfo {author} {\bibfnamefont {A.~H.}\ \bibnamefont
  {MacDonald}}\ and\ \bibinfo {author} {\bibfnamefont {U.}~\bibnamefont
  {Ekenberg}},\ }\href {\doibase 10.1103/PhysRevB.39.5959} {\bibfield
  {journal} {\bibinfo  {journal} {Phys. Rev. B}\ }\textbf {\bibinfo {volume}
  {39}},\ \bibinfo {pages} {5959} (\bibinfo {year} {1989})}\BibitemShut
  {NoStop}%
\bibitem [{\citenamefont {Yoshioka}(1986)}]{Yoshioka}%
  \BibitemOpen
  \bibfield  {author} {\bibinfo {author} {\bibfnamefont {D.}~\bibnamefont
  {Yoshioka}},\ }\href {\doibase 10.1143/JPSJ.55.885} {\bibfield  {journal}
  {\bibinfo  {journal} {J. Phys. Soc. Jpn.}\ }\textbf {\bibinfo {volume}
  {55}},\ \bibinfo {pages} {885} (\bibinfo {year} {1986})}\BibitemShut
  {NoStop}%
\bibitem [{\citenamefont {Kukushkin}\ \emph {et~al.}(2009)\citenamefont
  {Kukushkin}, \citenamefont {Smet}, \citenamefont {Scarola}, \citenamefont
  {Umansky},\ and\ \citenamefont {von Klitzing}}]{Kukushkin22052009}%
  \BibitemOpen
  \bibfield  {author} {\bibinfo {author} {\bibfnamefont {I.~V.}\ \bibnamefont
  {Kukushkin}}, \bibinfo {author} {\bibfnamefont {J.~H.}\ \bibnamefont {Smet}},
  \bibinfo {author} {\bibfnamefont {V.~W.}\ \bibnamefont {Scarola}}, \bibinfo
  {author} {\bibfnamefont {V.}~\bibnamefont {Umansky}}, \ and\ \bibinfo
  {author} {\bibfnamefont {K.}~\bibnamefont {von Klitzing}},\ }\href {\doibase
  10.1126/science.1171472} {\bibfield  {journal} {\bibinfo  {journal}
  {Science}\ }\textbf {\bibinfo {volume} {324}},\ \bibinfo {pages} {1044}
  (\bibinfo {year} {2009})}\BibitemShut {NoStop}%
\bibitem [{\citenamefont {Papi\ifmmode~\acute{c}\else \'{c}\fi{}}\ \emph
  {et~al.}(2012)\citenamefont {Papi\ifmmode~\acute{c}\else \'{c}\fi{}},
  \citenamefont {Haldane},\ and\ \citenamefont {Rezayi}}]{Papic-Haldane}%
  \BibitemOpen
  \bibfield  {author} {\bibinfo {author} {\bibfnamefont {Z.}~\bibnamefont
  {Papi\ifmmode~\acute{c}\else \'{c}\fi{}}}, \bibinfo {author} {\bibfnamefont
  {F.~D.~M.}\ \bibnamefont {Haldane}}, \ and\ \bibinfo {author} {\bibfnamefont
  {E.~H.}\ \bibnamefont {Rezayi}},\ }\href {\doibase
  10.1103/PhysRevLett.109.266806} {\bibfield  {journal} {\bibinfo  {journal}
  {Phys. Rev. Lett.}\ }\textbf {\bibinfo {volume} {109}},\ \bibinfo {pages}
  {266806} (\bibinfo {year} {2012})}\BibitemShut {NoStop}%
\bibitem [{\citenamefont {Read}\ and\ \citenamefont {Rezayi}(1996)}]{RRp0}%
  \BibitemOpen
  \bibfield  {author} {\bibinfo {author} {\bibfnamefont {N.}~\bibnamefont
  {Read}}\ and\ \bibinfo {author} {\bibfnamefont {E.}~\bibnamefont {Rezayi}},\
  }\href {\doibase 10.1103/PhysRevB.54.16864} {\bibfield  {journal} {\bibinfo
  {journal} {Phys. Rev. B}\ }\textbf {\bibinfo {volume} {54}},\ \bibinfo
  {pages} {16864} (\bibinfo {year} {1996})}\BibitemShut {NoStop}%
\bibitem [{\citenamefont {Levin}\ \emph {et~al.}(2007)\citenamefont {Levin},
  \citenamefont {Halperin},\ and\ \citenamefont {Rosenow}}]{aPf-Levin}%
  \BibitemOpen
  \bibfield  {author} {\bibinfo {author} {\bibfnamefont {M.}~\bibnamefont
  {Levin}}, \bibinfo {author} {\bibfnamefont {B.~I.}\ \bibnamefont {Halperin}},
  \ and\ \bibinfo {author} {\bibfnamefont {B.}~\bibnamefont {Rosenow}},\ }\href
  {\doibase 10.1103/PhysRevLett.99.236806} {\bibfield  {journal} {\bibinfo
  {journal} {Phys. Rev. Lett.}\ }\textbf {\bibinfo {volume} {99}},\ \bibinfo
  {pages} {236806} (\bibinfo {year} {2007})}\BibitemShut {NoStop}%
\bibitem [{\citenamefont {Lee}\ \emph {et~al.}(2007)\citenamefont {Lee},
  \citenamefont {Ryu}, \citenamefont {Nayak},\ and\ \citenamefont
  {Fisher}}]{aPf-Lee}%
  \BibitemOpen
  \bibfield  {author} {\bibinfo {author} {\bibfnamefont {S.-S.}\ \bibnamefont
  {Lee}}, \bibinfo {author} {\bibfnamefont {S.}~\bibnamefont {Ryu}}, \bibinfo
  {author} {\bibfnamefont {C.}~\bibnamefont {Nayak}}, \ and\ \bibinfo {author}
  {\bibfnamefont {M.~P.~A.}\ \bibnamefont {Fisher}},\ }\href {\doibase
  10.1103/PhysRevLett.99.236807} {\bibfield  {journal} {\bibinfo  {journal}
  {Phys. Rev. Lett.}\ }\textbf {\bibinfo {volume} {99}},\ \bibinfo {pages}
  {236807} (\bibinfo {year} {2007})}\BibitemShut {NoStop}%
\bibitem [{\citenamefont {Rezayi}\ and\ \citenamefont
  {Haldane}(2000)}]{Rezayi-Haldane}%
  \BibitemOpen
  \bibfield  {author} {\bibinfo {author} {\bibfnamefont {E.~H.}\ \bibnamefont
  {Rezayi}}\ and\ \bibinfo {author} {\bibfnamefont {F.~D.~M.}\ \bibnamefont
  {Haldane}},\ }\href {\doibase 10.1103/PhysRevLett.84.4685} {\bibfield
  {journal} {\bibinfo  {journal} {Phys. Rev. Lett.}\ }\textbf {\bibinfo
  {volume} {84}},\ \bibinfo {pages} {4685} (\bibinfo {year}
  {2000})}\BibitemShut {NoStop}%
\bibitem [{\citenamefont {Rezayi}\ and\ \citenamefont
  {Read}(1994)}]{Rezayi-Read-state}%
  \BibitemOpen
  \bibfield  {author} {\bibinfo {author} {\bibfnamefont {E.}~\bibnamefont
  {Rezayi}}\ and\ \bibinfo {author} {\bibfnamefont {N.}~\bibnamefont {Read}},\
  }\href {\doibase 10.1103/PhysRevLett.72.900} {\bibfield  {journal} {\bibinfo
  {journal} {Phys. Rev. Lett.}\ }\textbf {\bibinfo {volume} {72}},\ \bibinfo
  {pages} {900} (\bibinfo {year} {1994})}\BibitemShut {NoStop}%
\bibitem [{\citenamefont {Morf}(1998)}]{Morf}%
  \BibitemOpen
  \bibfield  {author} {\bibinfo {author} {\bibfnamefont {R.~H.}\ \bibnamefont
  {Morf}},\ }\href {\doibase 10.1103/PhysRevLett.80.1505} {\bibfield  {journal}
  {\bibinfo  {journal} {Phys. Rev. Lett.}\ }\textbf {\bibinfo {volume} {80}},\
  \bibinfo {pages} {1505} (\bibinfo {year} {1998})}\BibitemShut {NoStop}%
\bibitem [{\citenamefont {Papi\ifmmode~\acute{c}\else \'{c}\fi{}}\ \emph
  {et~al.}(2009)\citenamefont {Papi\ifmmode~\acute{c}\else \'{c}\fi{}},
  \citenamefont {Regnault},\ and\ \citenamefont
  {Das~Sarma}}]{LLLCoulombState-as-CFFS-p2}%
  \BibitemOpen
  \bibfield  {author} {\bibinfo {author} {\bibfnamefont {Z.}~\bibnamefont
  {Papi\ifmmode~\acute{c}\else \'{c}\fi{}}}, \bibinfo {author} {\bibfnamefont
  {N.}~\bibnamefont {Regnault}}, \ and\ \bibinfo {author} {\bibfnamefont
  {S.}~\bibnamefont {Das~Sarma}},\ }\href {\doibase 10.1103/PhysRevB.80.201303}
  {\bibfield  {journal} {\bibinfo  {journal} {Phys. Rev. B}\ }\textbf {\bibinfo
  {volume} {80}},\ \bibinfo {pages} {201303} (\bibinfo {year}
  {2009})}\BibitemShut {NoStop}%
\bibitem [{uqc()}]{uqc}%
  \BibitemOpen
  \href@noop {} {}\bibinfo {note} {M.~H. Freedman, A. Kitaev, M.~J. Larsen, and
  Z. Wang, arXiv:quant-ph/0101025 (unpublished)}\BibitemShut {NoStop}%
\bibitem [{\citenamefont {Haldane}(1983)}]{Haldane-Hierarchy}%
  \BibitemOpen
  \bibfield  {author} {\bibinfo {author} {\bibfnamefont {F.~D.~M.}\
  \bibnamefont {Haldane}},\ }\href {\doibase 10.1103/PhysRevLett.51.605}
  {\bibfield  {journal} {\bibinfo  {journal} {Phys. Rev. Lett.}\ }\textbf
  {\bibinfo {volume} {51}},\ \bibinfo {pages} {605} (\bibinfo {year}
  {1983})}\BibitemShut {NoStop}%
\bibitem [{\citenamefont {Halperin}(1984)}]{Halperin-Hierarchy}%
  \BibitemOpen
  \bibfield  {author} {\bibinfo {author} {\bibfnamefont {B.~I.}\ \bibnamefont
  {Halperin}},\ }\href {\doibase 10.1103/PhysRevLett.52.1583} {\bibfield
  {journal} {\bibinfo  {journal} {Phys. Rev. Lett.}\ }\textbf {\bibinfo
  {volume} {52}},\ \bibinfo {pages} {1583} (\bibinfo {year}
  {1984})}\BibitemShut {NoStop}%
\bibitem [{\citenamefont {Jain}(1989)}]{Jain}%
  \BibitemOpen
  \bibfield  {author} {\bibinfo {author} {\bibfnamefont {J.~K.}\ \bibnamefont
  {Jain}},\ }\href {\doibase 10.1103/PhysRevLett.63.199} {\bibfield  {journal}
  {\bibinfo  {journal} {Phys. Rev. Lett.}\ }\textbf {\bibinfo {volume} {63}},\
  \bibinfo {pages} {199} (\bibinfo {year} {1989})}\BibitemShut {NoStop}%
\end{thebibliography}
\end{document}